\documentclass[overfull]{epl}
\usepackage{graphicx}
\usepackage{cite}

\def\fsolv{F_{{\rm solv}}^{(m)}(T)}
\def\eintr{E_{{\rm intr}}^{(m)}}
\def\ftot {{\cal H}_{{\rm eff}}^{(m)}}
\def\erfc {\ {\rm erfc}}
\def\d    {{\rm d}}
\def\Q    {\langle Q \rangle}
\def\Nc   {\langle N_c \rangle}
\def\Tex  {T_{\rm ex}}
\def\Tw   {T_{\rm w}}
\def\Tc   {T_{\rm c}}
\def\Tcc  {T_{\rm d}}

\title{Four-states phase diagram of proteins.}

\author{Olivier Collet}
\institute{
\'equipe de Dynamique des Assemblages Membranaires, \\
UMR CNRS 7565, Facult\'{e} des Sciences, Universit\'{e} Henri Poincar\'{e}-Nancy 1, \\
54506 Vandoeuvre-l\`{e}s-Nancy, France}


\begin{document}

\maketitle

\begin{abstract}

A four states phase diagram for protein folding 
as a function of temperature and solvent quality
is derived from an improved 2-d lattice model taking into account the
temperature dependence of the hydrophobic effect.
The phase diagram exhibits native, globule and two coil-type regions. 
In agreement with experiment, the model reproduces the phase transitions indicative of 
both warm and cold denaturations.
Finally, it predicts transitions between the two coil states and
a critical point.

\end{abstract}

Understanding the physical mechanism underlying protein folding remains
one of the main open problems of contemporary theoretical biophysics.
The interplay between protein-protein and protein-solvent interactions
that drive the folding of the polypeptide may be partly investigated using
full atomistic representations. Computer simulations at this level of detail
are shown for instance to provide crucial information about the stability
of the proteins around its native structure. Such calculations are however
very time consuming and not appropriate for characterizing the large
conformational space of multimeric chains, which is a crucial step toward
understanding the folding problem.\cite{Karplus1992}.

This has led to the emergence of alternative approaches, such as
the use of simpler coarse grained models. Among these, the lattice
model is probably the most popular and efficient model that allows
a wide  sampling of the conformational space of a given
polypeptide chain \cite{Shakhnovich1990a}. 
Accordingly, the 16-mer placed
on a two dimensional lattice has often been used to this end.
\cite{Dinner1994, Cieplak1996, Collet2003b} Such a
chain is long enough to capture fundamental mechanism of protein
folding and short enough to allow the calculation of partition
function by a full enumeration in reasonable computer times.

Over a decade ago, Dinner et {\it al.}\cite{Dinner1994} used such a
model to derive the three-states phase diagrams of 16 mers for different
chain sequences as a function of temperature and average attraction between
monomers. Coil, globule and native states were all obtained but the
model failed to reproduce the well known cold denaturation. This transition,
from the native to the coil state, upon lowering the temperature, consists in
the loss of the order of the chain \cite{Privalov1990}.

It was indeed later shown that the accuracy of the potential describing
the interactions with the solvent is crucial\cite{Bryngelson1994}. We have
recently proposed a refinement of the coupling model that explicitly includes
a temperature dependent so-called hydrophobic effect in a solvation free energy 
contribution \cite{Collet2001}. 
This model, considering
the same 16-mer chain, predicts the existence of the coil and a native states,
the warm and cold denaturation transition but produces no globule states.
Despite this last shortcoming the coupling models were shown
to be consistent with all-atom molecular dynamics simulations of a short peptide
solvated in water\cite{Collet2003a}. In the recent litterature, 
other models dealing with the cold denaturation have also been proposed
\cite{DeLosRios2000}.

\begin{figure}[htb]
\centerline{\includegraphics[width=3.0cm]{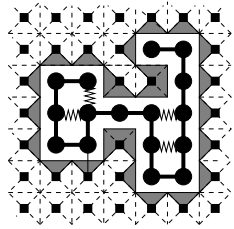}}
\caption{ \label{ex1} One conformation, of a 16 monomers chain (filled
circles) on a two-dimensional lattice. The thick solid lines
represent the covalent bonds and the springs the
intrachain contacts. The solvent sites are depicted as squares each of which
is divided into four solvent cells (triangles). Solvent-solvent interactions
involve two adjacent solvent cells (clear triangles)
whereas a solvent-monomer bond involves a monomer and a
nearest solvent cell (grey triangles).}
\end{figure}

In this paper, we extend on previous calculations and propose a more
comprehensive model of the hydrophobic effect that reproduces a four states phase
diagram with both the warm and cold denaturation transitions. 
In the model, all the links between two adjacent nods of the lattice
 are taken into account (see an example in fig.\ref{ex1}).
The effective hamiltonian of conformation $m$ is given by:

\begin{equation}
\label{eq3}
  \ftot(T) = \eintr + \fsolv
\end{equation}

The intrachain interaction energy for each conformation $m$
is described as in Dinner et {\it al.}\cite{Dinner1994}:
\begin{equation}
   \eintr = \sum_{i>j}^N B_{ij} \ \Delta_{ij}^{(m)} 
\end{equation}
where $B_{ij}$ is the specific interaction between residue sites
$i$ and $j$ and $\Delta_{ij}^{(m)}$ equals 1 if $i$ and $j$ are in
contact and 0 otherwise. Monomer-monomer interactions $B_{ij}$ are
real numbers selected randomly from  a normalized probability density
- Gaussian distribution -  with a standard deviation $\sigma = 2$.
One single conformation, noted Nat, is selected at random among the more
maximally compact structures, and considered as the native
conformation of the sequence. Nat has 9 intrachain contacts.
In the spirit of the Go-model\cite{Go1981}, the corresponding values
of interactions are described by the 9 smallest values of $B_{ij}$.

The free energy of solvation for each conformation may be written as
a sum of two contributions: 
\begin{equation}
F_{\rm solv}^{(m)}(T) = \sum_{i=1}^N n_i^{(m)} f_i(T) + 2 n_s^{(m)} f_s(T)
\end{equation}
Where $n_i^{(m)}$ and $n_s^{(m)}$ are respectively the number of solvent
sites surrounding residue $i$ and the total number of solvent-solvent
contacts. $f_{i}(T)$ is the specific free energy of a
solvent cell in interaction with residue $i$ and $f_s(T)$, that of a neat
solvent cell.

{

Taken the extended structures (without any intrachain contacts) as 
the free energy reference, the effective hamiltonian may be rewritten
as a summation of effective couplings between monomers
(see the example of figure \ref{bijeff})~:
\begin{equation}
\ftot(T) = 
\sum_{i>j}^N B_{ij}^{\rm eff}(T) \ \Delta_{ij}^{(m)} \quad {\rm with} \quad
B_{ij}^{\rm eff}(T) = B_{ij} - f_i(T) - f_j(T) + 2 f_s(T)
\end{equation}
\begin{figure}[hb]
\centerline{
\includegraphics[width=3.0cm]{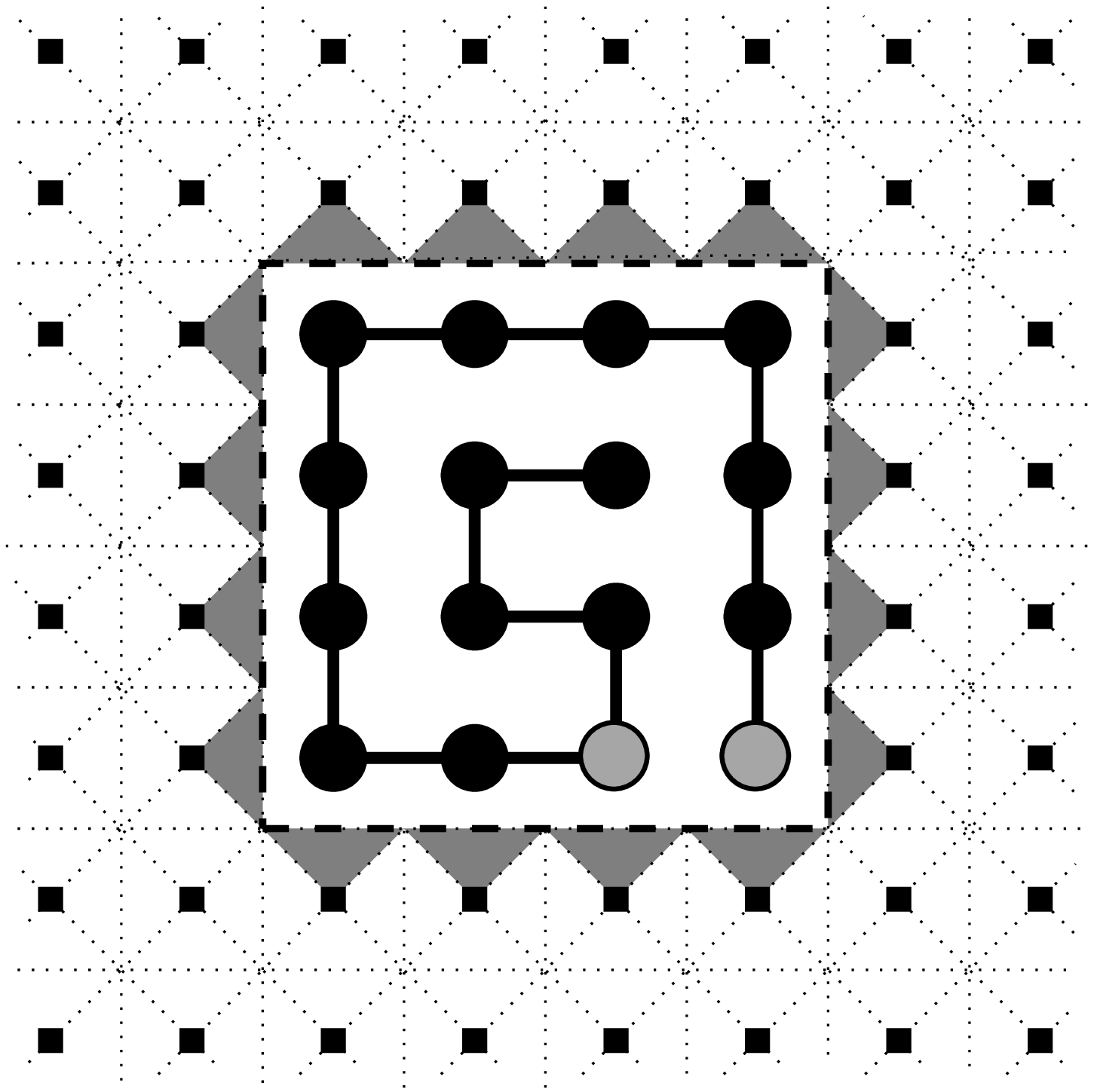}
\hskip+1.6cm
\includegraphics[width=3.0cm]{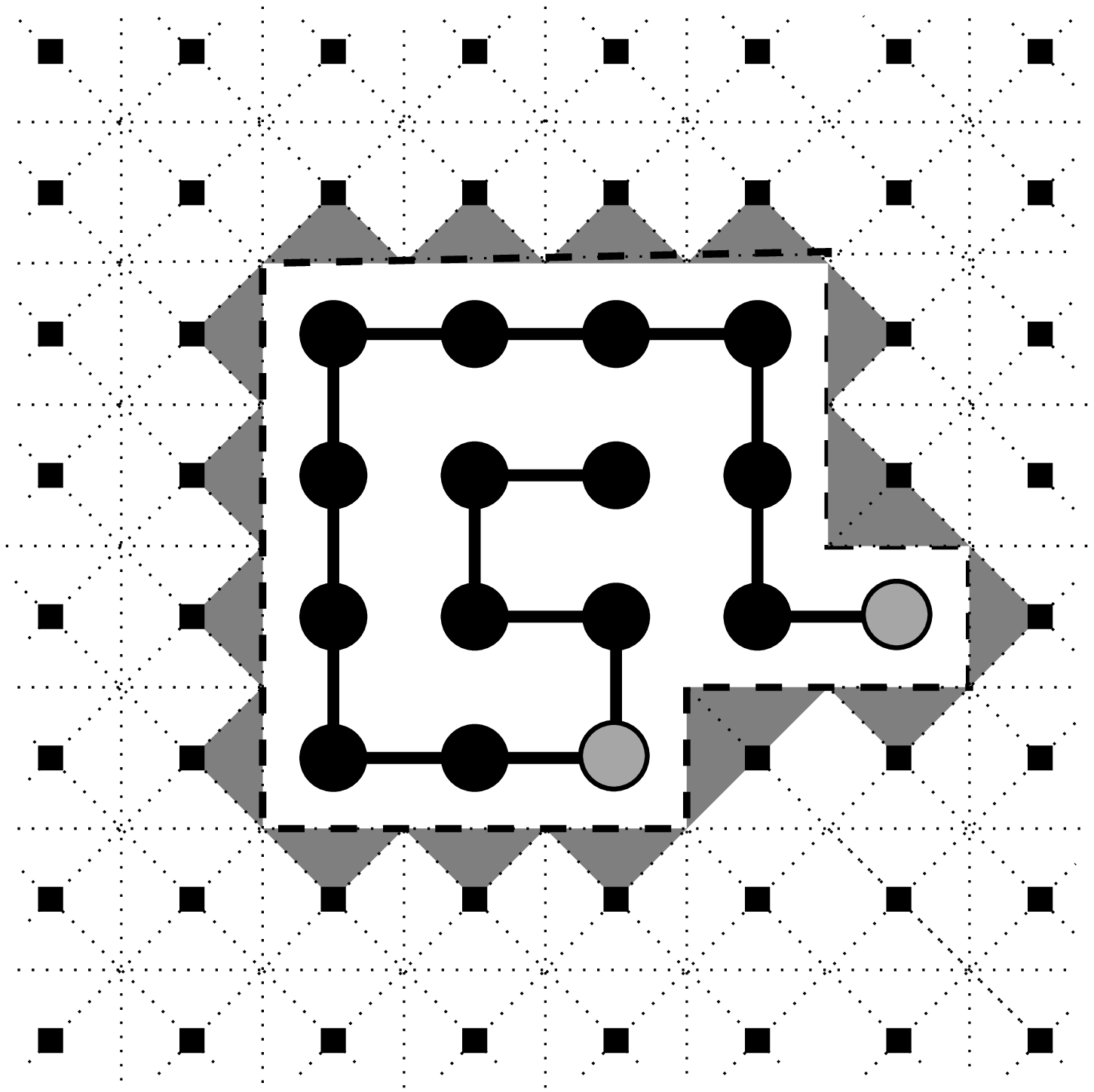}}
\vskip 0cm
\caption{\label{bijeff}
Structure $A$ and $B$ differs only by the contact between monomers 5 and 16.
The effective couplings between these monomers is simply the effective
hamiltonian difference between conformations $A$ and $B$, calculated by a counting
of the lattice links~:
$B_{5,16}^{\rm eff} = B_{5,16} - f_5(T) - f_{16}(T) + 2 f_s(B_s, T)$
}
\end{figure}

}

{

Recently, Silverstein et {\it al.}\cite{Silverstein1999} gave a description of 
the hydrophobic effect in terms of two energy spectra that best fits their
simulation data. These results exhibits a low degenerated, narrow, 
(respectively high degenerated, extended,) spectra for neat water
(respectively for aqueous solution with a non polar solute).
Here, this physical picture is reduced further.}
The energy spectra of the solvent in interaction with monomer
$i$ consists of $N_s$ energy values $B_i^{(j)}$, ($j=1,N_s$) selected from a
Gaussian distribution with standard deviation $\sigma$, while the energy spectrum
of the neat solvent is given by a unique level, $N_s^\alpha$-fold degenerated,
of energy $B_s$. Small values of $B_s$ models bad solvent and large values
good solvent. 
Extending on our previous model, \cite{Collet2001} we introduce here an
extra parameter $\alpha$, representing the degeneracy ratio between
the bulk and the first shell solvent cells. 
As the total degeneracy of the latter is higher than that of the former
\cite{Silverstein1999},
one has $\alpha < 1$ and these degeneracies being related to the number of
solvent configurations $N_s$ is a large number.

Accordingly, the free energies associated
with the neat solvent and that of solvation of each monomer $i$ are
respectively given by~:
\begin{eqnarray}
f_s(B_s,T) & = & B_s - \alpha T \ln N_s \\
f_i(T)     & = & -T \ln z_i(T)
\end{eqnarray}
where $z_i(T)$ is the partition function of the solvent around monomer $i$.
For large values of $N_s$, it may be written using a continuous formalism as~:
\begin{equation}
z_i(T)  =  N_s \int_{B_i^{\rm min}}^\infty n(B_i) \exp\left(
- \frac{B_i}{T} \right) \d B_i
\end{equation}
where $n(B_i)$ is the normalized Gaussian distribution truncated at
$B_i^{\rm min} = \min_j B_i^{(j)}$, specific to each residue:
\begin{equation}
n(B_i) =
\left\{ \begin{array}{lll}
0 & \ \ {\rm if} \ \ & B <  B_i^{\rm min} \\
\frac
{\exp\left(-\frac{B_i^2}{2 \sigma^2} \right) }
{\frac{\sigma}{2} \sqrt {2 \pi} \erfc
 \left(\frac{B_i^{\rm min}}{\sigma \sqrt 2} \right)}
& \ \ {\rm if} \ \  & B \ge B_i^{\rm min}
\end{array} \right.
\end{equation}
Equation (6) may therefore be rewritten as:
\begin{equation}
z_i(T) =  N_s \exp \left(\frac{\sigma^2}{2 T^2} \right)
\frac{\erfc \left(\frac{B_i^{\rm min}}{\sigma \sqrt 2} +
    \frac{\sigma \sqrt 2}{2T}\right)}
     {\erfc \left(\frac{B_i^{\rm min}}{\sigma \sqrt 2}\right)}
\end{equation}

The density of probability that the smallest value of the $N_s$ set,
chosen at random with a Gaussian distribution, be $B^{\rm min}$, is~:
\begin{equation}
{\cal G}_{N_s}(B^{\rm min}) = N_s \ g(B^{\rm min}) \ {\cal P} (x \ge B^{\rm min})^{N_s-1}
\end{equation}
where $g(B^{\rm min})$ is the density of probability to select $B^{\rm min}$
and ${\cal P}(x \ge B^{\rm min})$ the probability to draw a value $x$ larger than $B$.
Thus, for each residue $i$, $B_i^{\rm min}$ is selected from the probability
density:
\begin{equation}
{\cal G}_{N_s}(B) = \frac{N_s}{\sigma \sqrt{2 \pi}}
\exp\left(-\frac{B^2}{2 \sigma^2} \right)
\left( \frac{1}{2} \erfc \left(\frac{B}
{\sigma \sqrt 2} \right) \right)^{N_s-1}
\end{equation}

The state of the chain under each set of conditions is determined
from statistical equilibrium averages. For an observable $X^{(m)}$,
the average over peptide structures may be defined as:
\begin{equation}
\langle X(B_s,T) \rangle = 
\sum_{m=1}^\Omega X^{(m)} P_{eq}^{(m)}(B_s,T) 
\end{equation}
with 
\begin{equation}
P_{eq}(B_s,T) = \frac{\exp\left(-\frac{\ftot}{T} \right)}
{\sum_{m=1}^\Omega \exp(-\frac{\ftot}{T})}
\end{equation}

This expression allows the estimate of the chain entropy,
$ S_{ch}(B_s,T) = - \langle \ln P_{eq} \rangle $,
the compactness of the peptide, defined as the average of
$   N_c^{(m)}= \frac{1}{9}\sum_{i>j}^{N} \Delta_{ij}^{(m)} $ where
$\sum_{i>j}^{N} \Delta_{ij}^{(m)} $ is the number of intra-chain
contacts of structure $m$, and the order of the peptide, defined as the average
of $ Q^{(m)}$, the  pairwise contact overlap of the structures with
the native conformation $ \left( Q^{(m)}= \frac{1}{9}.
     \sum_{i>j}^{N} \Delta_{ij}^{(m)}\Delta_{ij}^{\rm Nat} \right)$.
The number of contacts of the more maximally compact structures (i.e. 9 for
the specific chain length studied here), appears in the two above averages
in order to normalized them to 1.

\begin{figure}[htb]
\centerline{
\includegraphics[width=8.5cm]{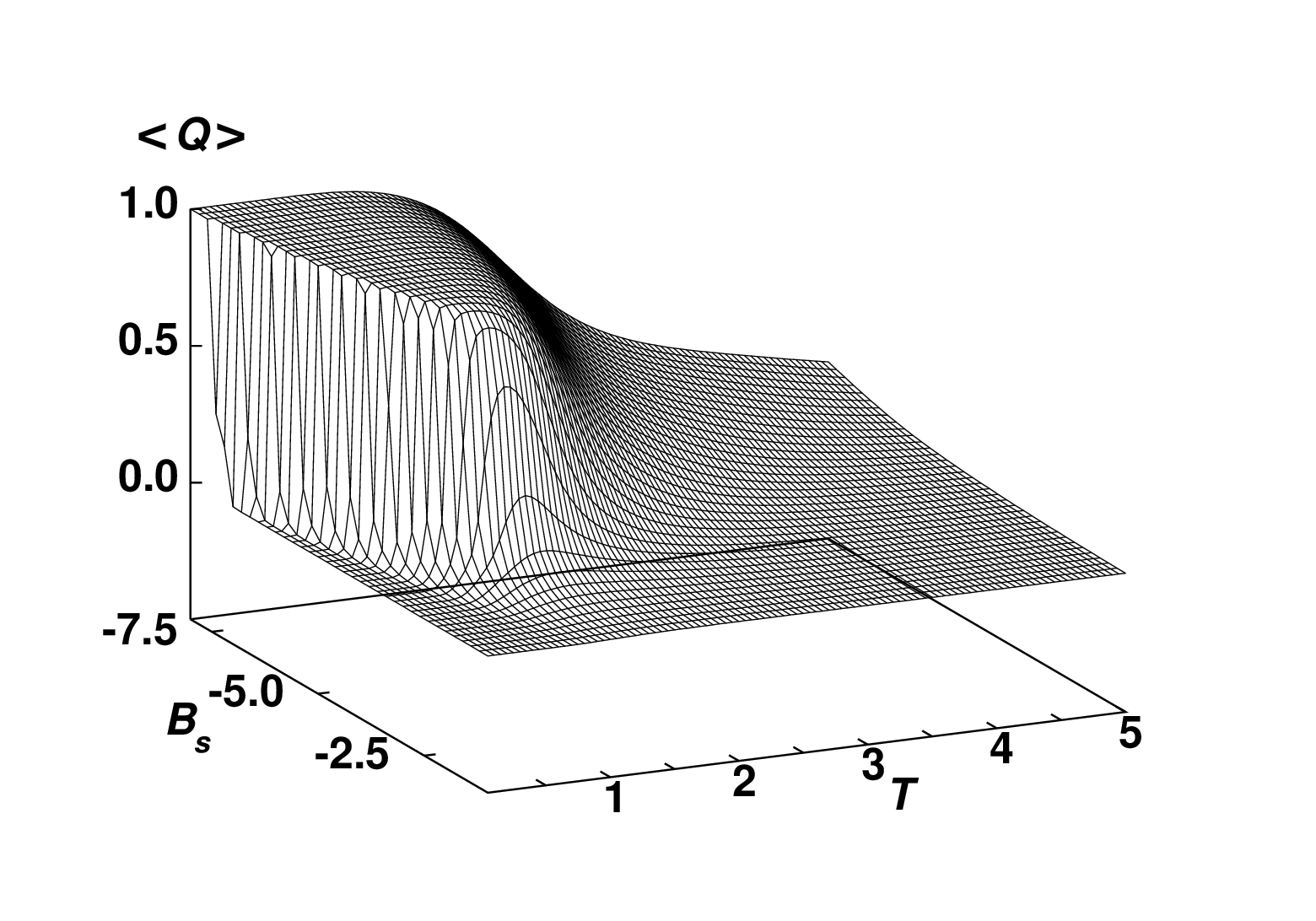}
\hskip-1.6cm
\includegraphics[width=8.5cm]{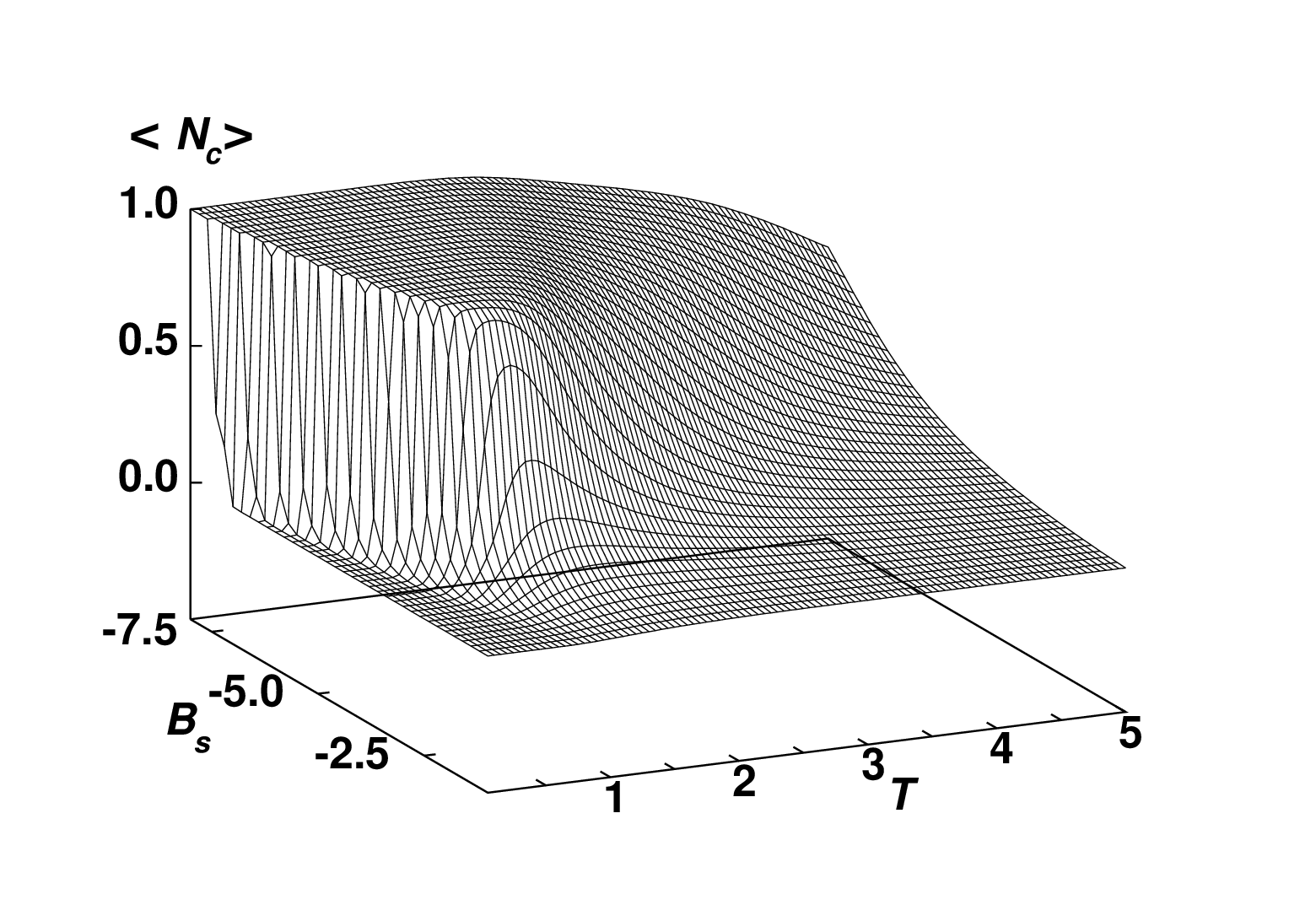}}
\vskip-1cm
\centerline{
\includegraphics[width=8.5cm]{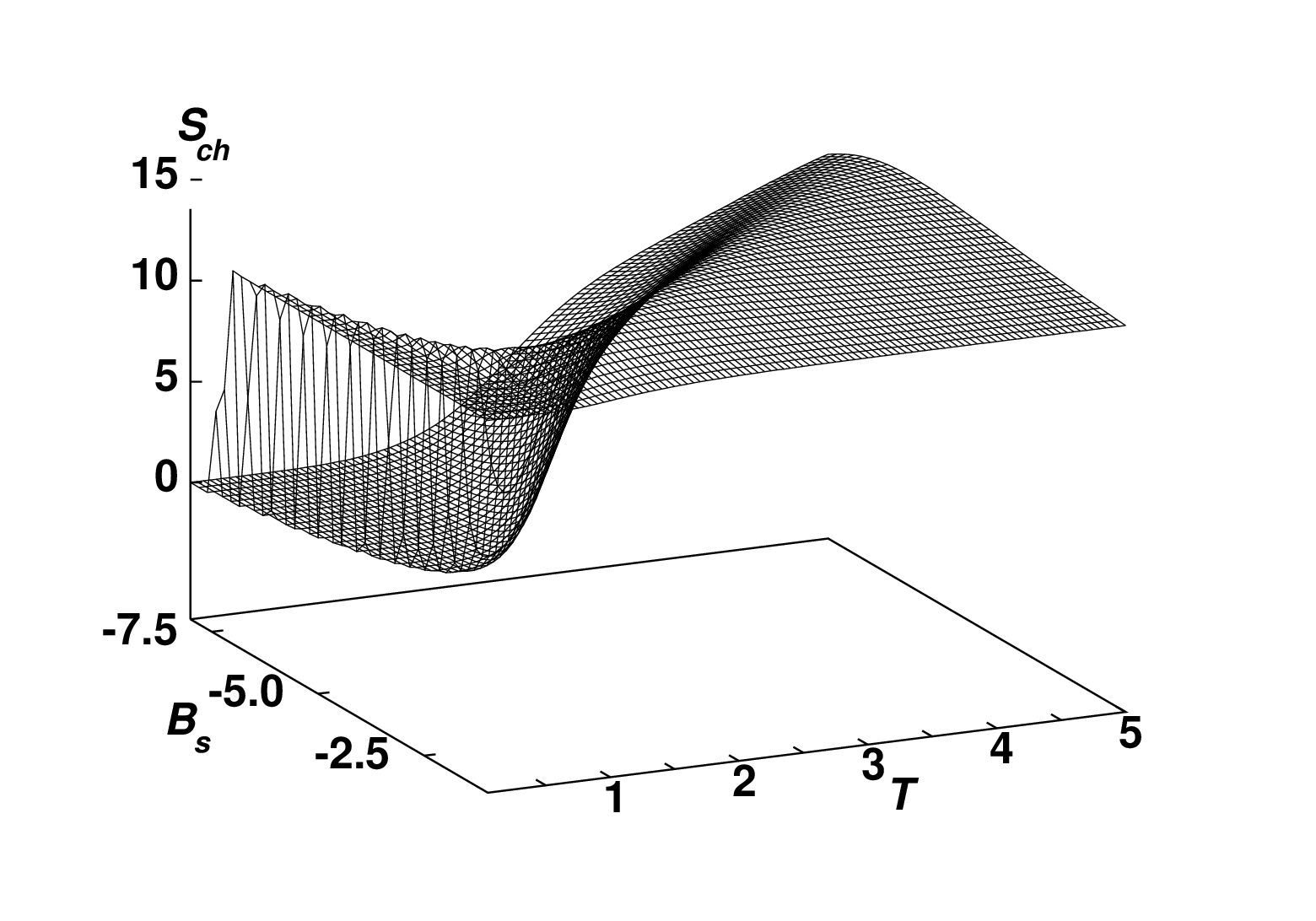}}
\vskip-1cm
\caption{\label{res}
Averages of the order parameter $\langle Q \rangle$, 
the compactness $\langle N_s \rangle$ and
the chain entropy $S_{\rm ch}$ as functions of $B_s$ and $T$.
}
\end{figure}

For a 16-mer chain model on a two dimensional lattice,
the total number of structures is $n_{\rm tot} = 802 075$
among which $n_{\rm ext} = 116 579$ have zero contact.
For given values of $B_s$ and $T$, the point state in the phase diagram
is determined by $\Q$, $\Nc$ and $S_{\rm ch}$. When $\Q > 0.66$, the peptide is
considered in the Native phase. When $\Q < 0.66$  and $\Nc > 0.66$, only
some compact structures are relevant, and the chain is in the so-called
globule state. When $\Nc < 0.66$ and $S_{\rm ch} = \ln n_{\rm ext}$,
the peptide is mainly in the extended conformation and the phase is coil type II.
Last, when $\Nc < 0.66$ and $S_{\rm ch} > \ln n_{\rm ext}$, almost all chain
structures have a non zero probability to occur. This state is referred to as
coil-type I.

By setting the model parameters to $\sigma =0$ and $\alpha=1$, the temperature
dependence of the hydrophobic effect is effectively removed. Under such conditions,
the corresponding phase diagrams is similar to that determined by
Dinner et {\it al.}\cite{Dinner1994}.
On the other hand, 
the two states phase diagram where the warm and cold denaturation are present
\cite{Collet2001} may be obtained by setting $\sigma=2$, $\alpha=0.5$ and $N_s=10^5$.

For discussing the four state phase diagram, we set, in the following, 
the model parameters to $\sigma=2$,  $\alpha=0.9$ and $N_s=10^5$. 
The results are mildly sequence dependant.
We therefore select a particular sequence, and investigate its corresponding
phase behavior. $\Nc$, $\Q$ and $S_{\rm ch}$ are displayed in fig.\ref{res}
as a function of $B_s$ and $T$. 
Several qualitative features are directly
observed from the 3-d plots. These may be classified depending on $B_s$
as follow:

For $B_s < -7.5 $, the $\Q$ plots indicate that the peptide is in the native phase
at low temperature and in denatured phase at high temperature. Depending on the
$B_s$ value, the transitions in $\Q$ and $\Nc$ take place at different temperatures,
noted hereafter $\Tw$ and $\Tex$ respectively 
(i.e. $\Q(B_s, \Tw(B_s)) = 0.66$ and $\Nc(B_s, \Tex(B_s))=0.66$).
As for $T>\Tex$, the chain entropy
becomes an increasing function of temperature (up to $\ln n_{\rm tot}$), one may
identify three regions corresponding to the following phases: a coil type I phase
for temperatures above $\Tex$, a globule phase between $\Tw$ and $\Tex$,
and a native state below $\Tw$.

For $-7.5  < B_s < -2.5$, in addition to the states described above,
transitions toward denatured states ($\langle Q \rangle \rightarrow 0$)
take place at low temperatures. Such transitions, occurring at temperatures
$\Tc$ that depend on $B_s$, represent cold denaturation. Below $\Tc$, the chain
entropy is constant and equals $\ln n_{\rm ext}$ which indicates
that the low temperature region corresponds to the coil type II state.

\begin{figure}[htb]
{\includegraphics[width=7.5cm]{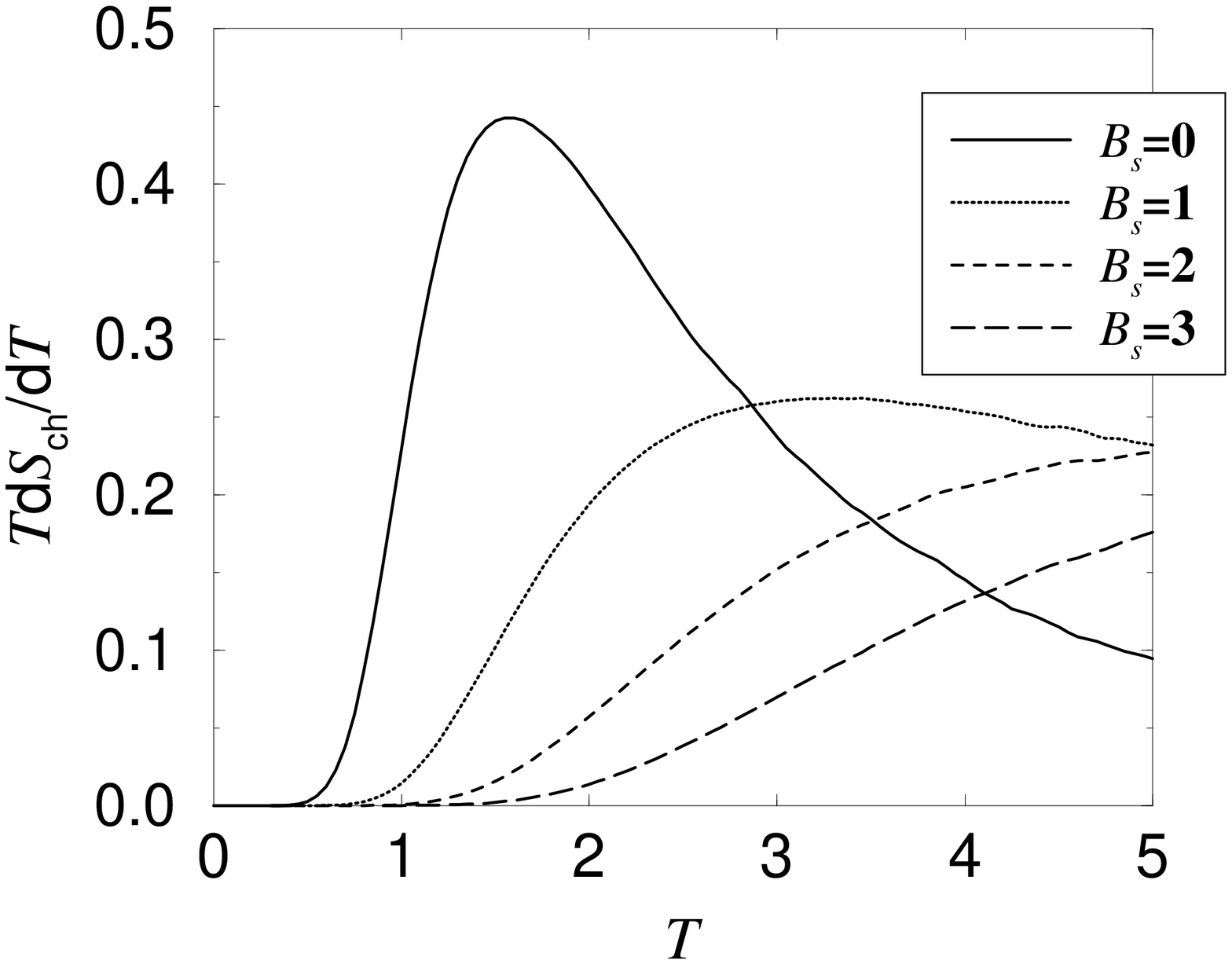}}
{
\hskip2cm  \includegraphics[height=3.0cm]{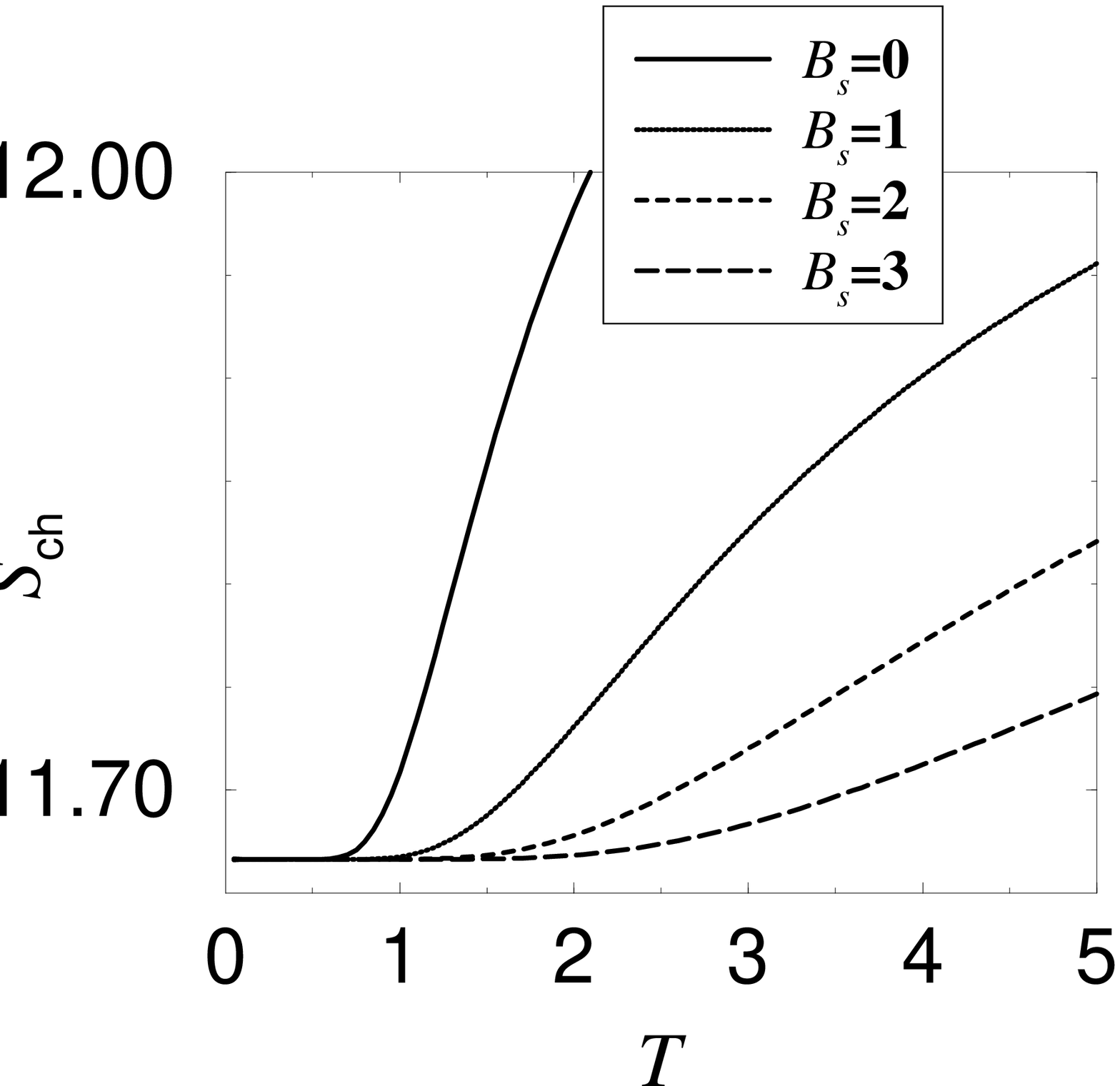}
\vskip-6.5cm\hskip9.4cm\includegraphics[height=3.1cm]{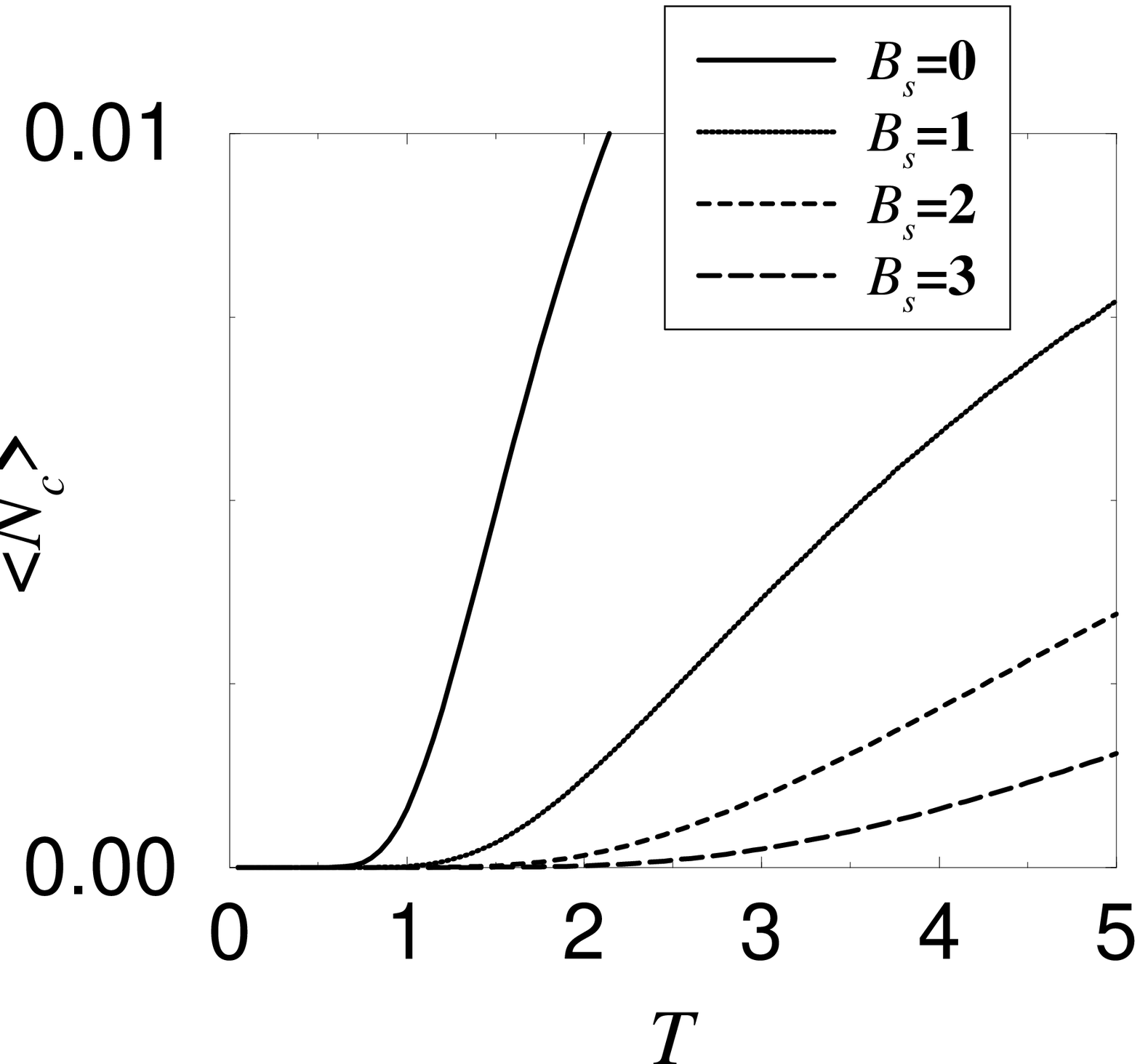} }
\vskip3.cm
\caption{
\label{Tdd}
Statistical averages of the chain in interaction with good solvents
modelled by $B_s = \{0, 1, 2, 3\}$ versus $T$.
Top-left: chain entropy.
Top-right: compactness.
Bottom: heat capacity.
}
\end{figure}

For $B_s$ values above -2.5, $\Q$, $\Nc$ are very small.
The chain is always in a coil state, regardless of the temperature.
These values of $B_s$ are therefore indicative of good solvation.
Different states are however observed as shown from the $\Nc$ and
$S_{\rm ch}$ plots (fig.\ref{Tdd}). 
At low temperature, the compactness is
rigourously null and the chain entropy equals $\ln n_{\rm ext}$,
indicating that the peptide is in coil type II state.
As the temperature increases, so does the entropy until reaching
$\ln n_{\rm tot}$ and the chain is in coil type I phase.
To better delineate the frontier between the coil type I and coil type II
regions, we have estimated numerically $T \frac{\d S_{\rm ch}}{\d T}$, the contribution
of the chain to the heat capacity of the system as a function of temperature.
For $B_s<2.0$, these contributions undergo a maxima at $T=\Tcc$, which is a signature
of a first order disordered-disordered transition between the two coil phases.
For $B_s>2.0$, the peak is no longer observed.

\begin{figure}[hbt]
\centerline{\includegraphics[width=8.5cm]{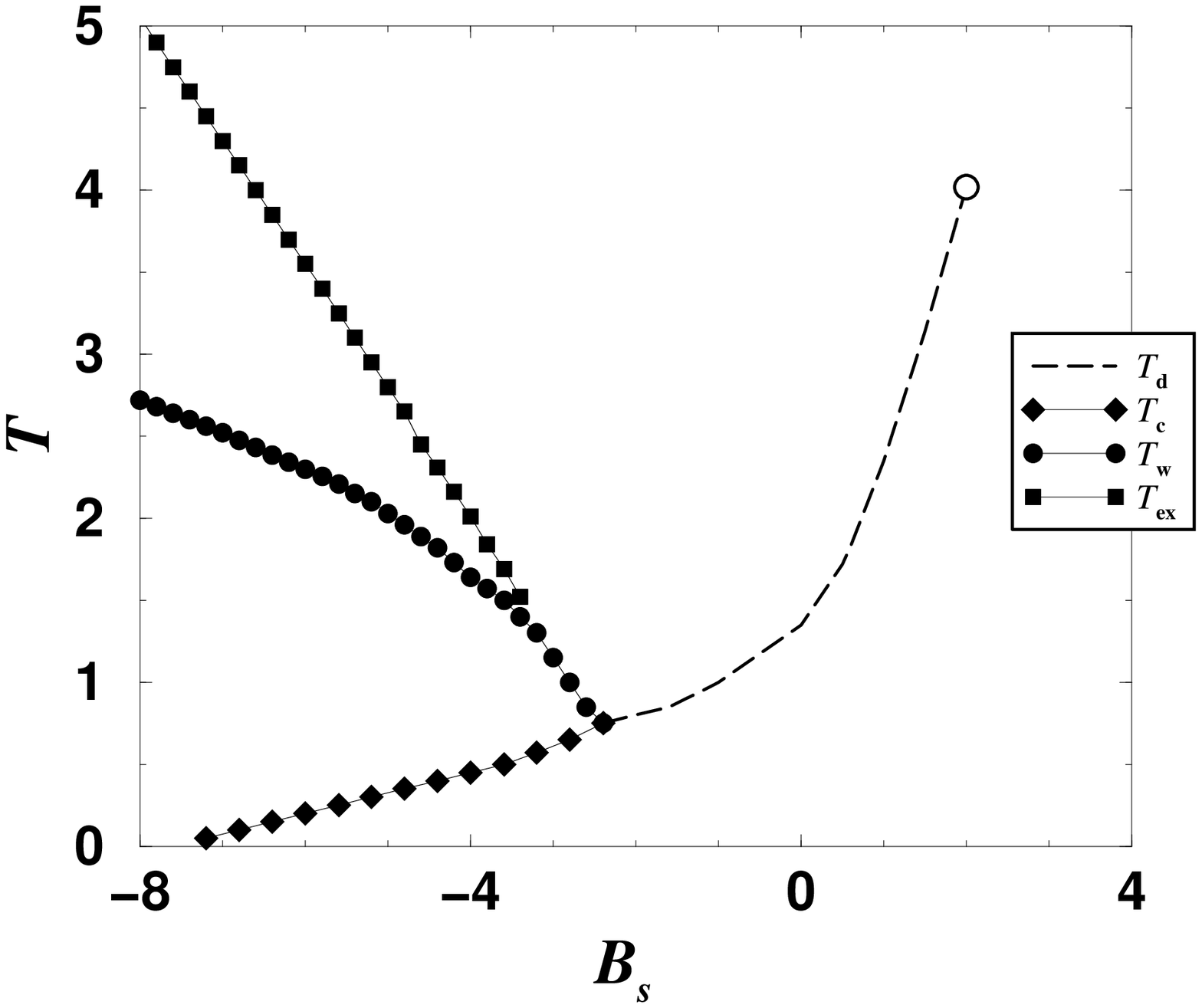}}
\vskip-3.3cm \hskip+5.0cm\includegraphics[width=0.6cm]{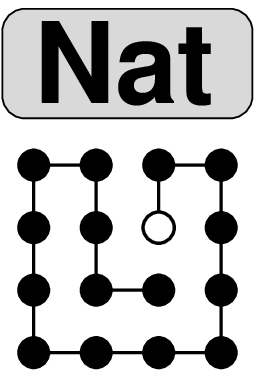}
\vskip-3.8cm \hskip+4.0cm\includegraphics[width=1.6cm]{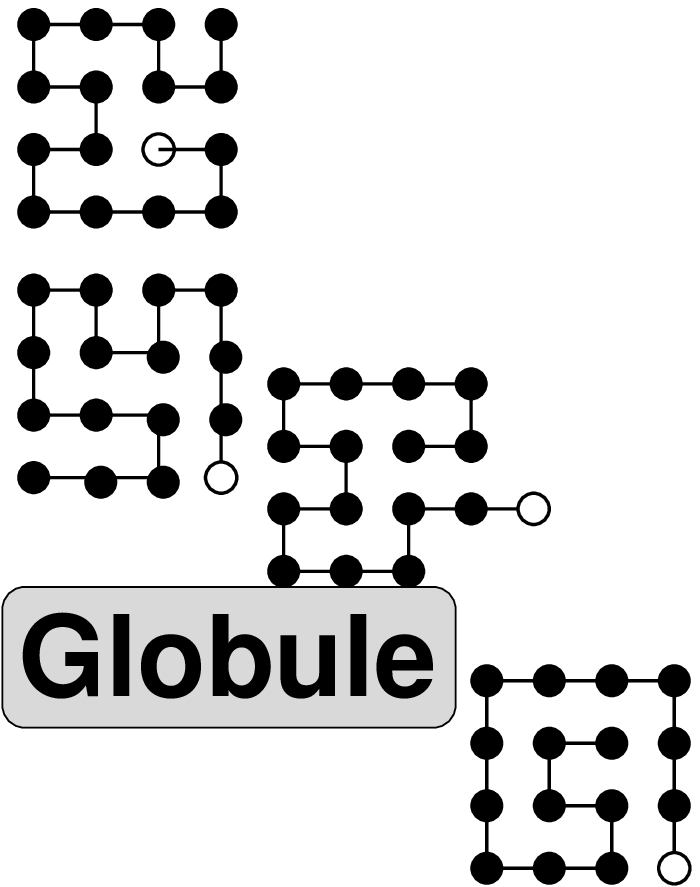}
\vskip-2.7cm \hskip+6.2cm\includegraphics[width=2.0cm]{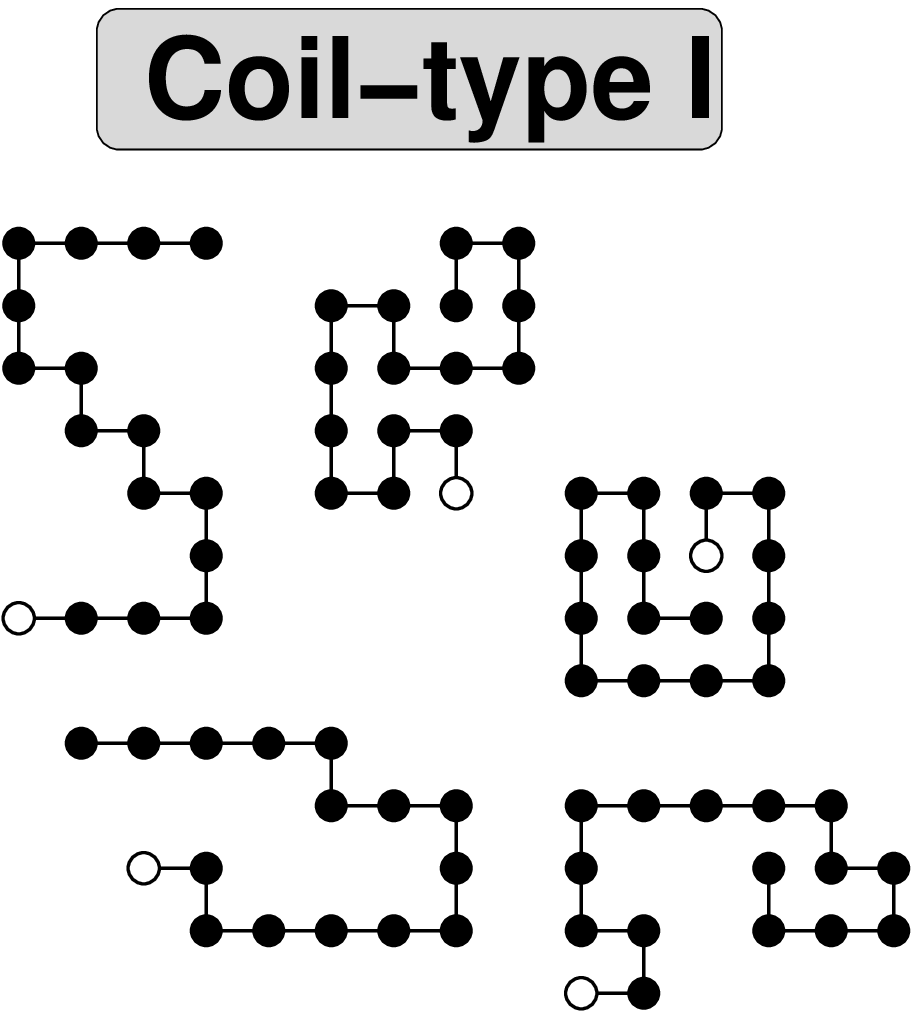}
\vskip+2.0cm \hskip+7.0cm \includegraphics[width=3.4cm]{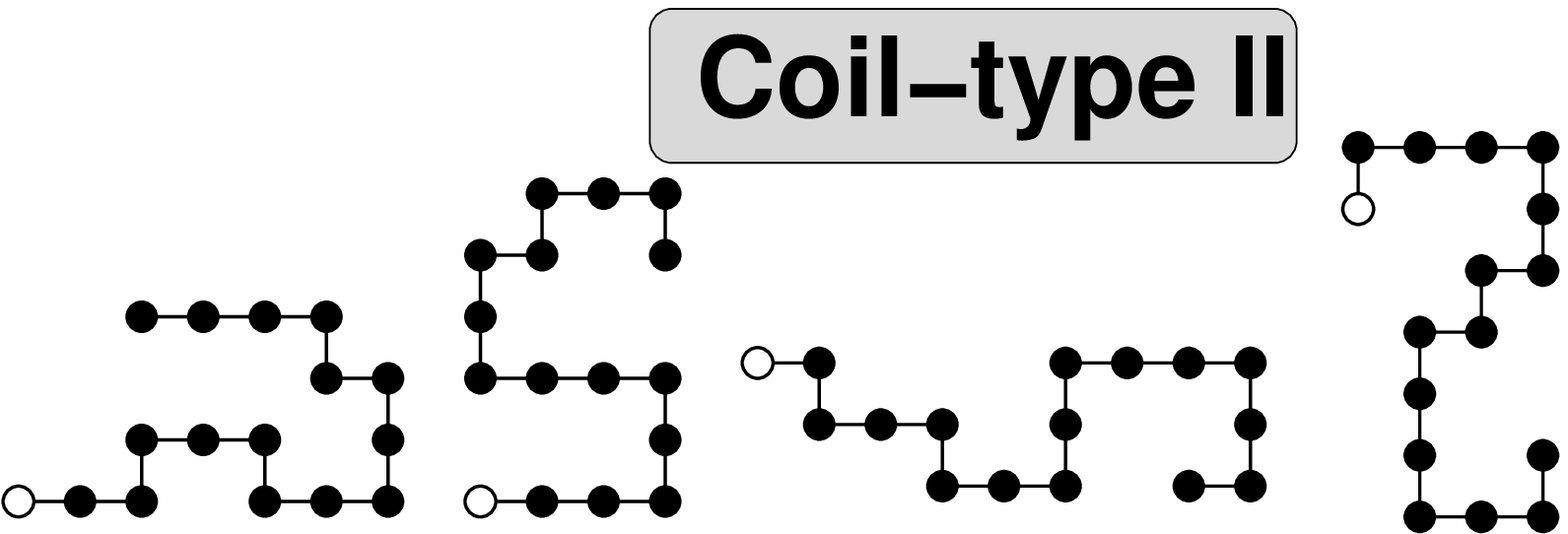}
\vskip+1.0cm
\caption{\label{pd2}
Phase diagram of the chain as a function of the solvent quality and the temperature.
All the structures are observed in the coil type I,
only the extended ones in the coil type II,
only the compact ones in the globule and
the sole native conformation in the nat region.
}
\end{figure}
The previous results are summarized in the phase diagram reported in fig.\ref{pd2}.
For the particular sequence considered here, four states are distinguished.
The native, globule and coil-type I phases coexist at the triple point:
$(B_s, T)=(-3.4, 1.40)$ and the native, coil-type I and
coil-type II phases coexist at $(B_s, T)=(-2.4, 0.72)$. 
A critical point is observed at $(B_s, T)_c = (2.0, 4.0)$.
{
Thus, moving along the $B_s$ and $T$ axes, transitions from Coil Type I and
Coil Type II without crossing any peak in the heat capacity are allowed.
Very small and smooth variations in $S_{\rm ch}$, $\Q$ and $\Nc$ occur on
these ways.
This confirms that Coil Type I and Coil Type II are two phases of the extended
state, which implies that warm and cold denaturations are, indeed, transitions
toward the same extented state.
}
The existence of a hypothetical supercritical phase for $B_s > 2.0$ or
$T> 4.0$ is not clear. The nature of the set of structures relevant in
such an speculative region should be investigated by the detailed study of
effective hamiltonian spectra as function of the temperature. 

{
Last, in the simulations performed with $\alpha = 0.5$\cite{Collet2001},
the Globule and Coil Type I phases disappear leaving only the
warm and the cold transitions between Coil Type II and Nat.}

In summary, we have shown that the suitable solvation model presented in this
paper allows to calculate, for the first time, a four-state phase diagram of
a peptide chain.
One would need, however, to elucidate the physical meaning of all the 
model parameters $N_s$, $\sigma$ and $\alpha$ and their relative values 
for the 20 natural amino acids for
a complete understanding of the mechanism responsible for protein folding.
Last, it must be understood that similar phase diagrams are obtained
if the same value of $B_i^{\rm min}$ is affected to every residue.
However, we choose to select one value of $B_i^{\rm min}$ 
for each residue in order to model the
specifity of the hydrophobicity of each monomer of the protein.

\acknowledgments
It is a pleasure to acknowledge Mounir Tarek for helpful discussions and critical reading
of the manuscript.

\end{document}